\documentclass[aip,jcp,amsmath,amssymb,floatfix,citeautoscript,reprint]{revtex4-1}

\usepackage{graphicx,graphics,epsfig,subfigure,times,bm,bbm,amssymb,amsmath,amsfonts,amsthm,mathrsfs,MnSymbol}
\usepackage[matrix,frame,arrow]{xypic}
\usepackage[normalem]{ulem}
\usepackage{slashed}
\usepackage{dcolumn}
\usepackage{tabularx}
\usepackage{color}
\usepackage[usenames,dvipsnames,svgnames,table]{xcolor}
\usepackage[english]{babel}
\usepackage{verbatim}
\definecolor{darkblue}{rgb}{0.0,0.0,0.3}
\usepackage[colorlinks=true,
            linkcolor=red,
            urlcolor= darkblue,
            citecolor=blue]{hyperref}
\newcommand{\bea}{\begin{eqnarray}}
\newcommand{\eea}{\end{eqnarray}}

\begin{document}

\title{Electron transfer at electrode interfaces via a straightforward quasiclassical fermionic mapping approach}

\author{Kenneth A. Jung}
\author{Joseph Kelly}
\author{Thomas E. Markland}
\email{tmarkland@stanford.edu}
\affiliation{Department of Chemistry, Stanford University, Stanford, California, 94305, USA}

\date{\today}

\begin{abstract}
Electron transfer at electrode interfaces to molecules in solution or at the electrode surface plays a vital role in numerous technological processes. However, treating these processes requires a unified and accurate treatment of the fermionic states of the electrode and their coupling to the molecule being oxidized or reduced in the electrochemical processes and, in turn, the way the molecular energy levels are modulated by the bosonic nuclear modes of the molecule and solvent. Here we present a physically transparent quasiclassical scheme to treat these electrochemical electron transfer processes in the presence of molecular vibrations by using an appropriately chosen mapping of the fermionic variables. We demonstrate that this approach, which is exact in the limit of non-interacting fermions, is able to accurately capture the electron transfer dynamics from the electrode even when the process is coupled to vibrational motions in regimes of weak coupling. This approach thus provides a scalable strategy to explicitly treat electron transfer from electrode interfaces in condensed-phase molecular systems.
\end{abstract}

\maketitle

\section{Introduction}

Reactions at electrode interfaces are ubiquitous in industrial processes and omnipresent in chemical, medical and energy research. Accurately capturing these processes requires treating the quantum dynamics of the electrode states and their coupling to the electronic and nuclear dynamics of the molecules in the electrolyte. The vastly different timescales of these dynamical processes as well as the combined size of the electronic and nuclear Hilbert spaces makes accurate simulations of these systems with atomisitic detail highly challenging. 

Semiclassical and quasiclassical methods, which use classical trajectories to approximate exact quantum dynamics, provide a potentially appealing route to treat the dynamics of condensed phase systems with lead(electrode)-molecule interactions due to their efficiency and low scaling with dimensionality. The two approaches that have generated the most methods for treating these systems semiclassically are those based on surface hopping and those that arise from mapping Hamiltonians.

Recently introduced surface hopping based methods include the independent electron surface hopping\cite{Shenvi2009,Shenvi2009ii,Gardner2023} (IESH) and the classical master equation evaluated with surface hopping\cite{Dou2015,Dou2015iii,Dou2016} (CME-SH) approaches. The CME-SH approach has been combined with a diffusive description of the surrounding condensed phase environment to describe a range of electrochemical properties.\cite{Coffman2018,Coffman2019,Coffman2020} The CME also provides a link to methods which incorporate the effects of the electrons in the lead using electronic friction\cite{HeadGordon1995,Dou2017,dou2018modeling}. However, the derivation of the CME assumes weak lead-molecule coupling and requires introducing an {\it ad hoc} broadening procedure to properly treat the lead-molecule hybridization for stronger coupling regimes. 

Semiclassical mapping methods, which map the discrete, single-particle creation and annihilation operators to continuous, classical degrees of freedom, provide an appealing approach to obtain an efficient classical-like description of the dynamics of lead-molecule systems. The Li-Miller mapping\cite{Li2012} (LMM) has been shown to give accurate results when applied to describing electronic transport through molecular junctions\cite{Li2013,Li2014,Levy2019} (the Anderson impurity model). The apparent success of LMM for these systems occurs despite its mapping of fermionic creation annihilation operators, which obey fermionic symmetries, onto Cartesian degrees of freedom that do not obey the fermionic anti-commutation identities. The more recent complete quasiclassical mapping\cite{Levy2019} (CQM) has extended the LMM to capture more general classes of fermionic observables. The success of the LMM and CQM approaches in accurately reproducing the dynamics in benchmark systems of nanoscopic transport despite not satisfying the fundamental anticommutivity of the individual fermionic operators initially seems quite remarkable and that a deeper reason exists for their success. Recent work has provided general insights into the reasons that mapping the fermionic dynamics of the lead onto bosons in many cases can give accurate or even exact results\cite{Sun2021,MontoyaCastillo2023}. In particular, it has been shown that, in the case of non-interacting electrons, one can exchange the fermionic and bosonic dynamics, while retaining the fermionic statistics, and yet still obtain the exact quantum dynamics\cite{Sun2021}. More generally the conditions (i.e. combination of the form of the Hamiltonian and observables) under which individual fermionic creation and annihilation operators can be rigorously replaced with their bosonic counterparts has been derived\cite{MontoyaCastillo2023}. 

Given the realization that under particular conditions one can obtain the exact dynamics by mapping the fermionic lead to bosons thus allows for the use of the Meyer-Miller (MM) mapping\cite{Meyer1980}, which has long been appreciated as an exact mapping in the case of bosons\cite{Stock1997}, to be applied to develop a classical-like description of fermionic processes. While Meyer-Miller mapping of the fermionic operators is no longer exact in the case of interacting fermions, recent work has shown that it can still provide an accurate description of purely fermionic systems.\cite{Sun2021} Here we investigate and analyze the accuracy of MM mapping in treating systems involving a mixture of fermionic (lead and molecule electronic occupations) and bosonic (nuclear motion) degrees of freedom. In particular, we investigate an Anderson-Holstein model consisting of a lead (electrode) that can transfer electrons to a molecule in the electrolyte with the transfer mediated by a vibrational mode. This Hamiltonian has formed the basis of numerous previous studies to benchmark techniques for use in electrochemical simulations \cite{Navrotskaya2009,Dou2015,Coffman2019,Coffman2020}. From this we are able to contrast the benefits of MM mapping to surface hopping based approaches. We show that the mapping approach has advantages over surface hopping approaches in that the states of the lead are straightforward to discretize and it is accurate for strong lead-molecule coupling. However, the mapping approach suffers from detailed balance issues that become increasingly apparent in the case of strong molecule-vibration coupling.

\section{Theory}

\begin{figure}[h]
    \begin{center}
        \includegraphics[width=0.45\textwidth]{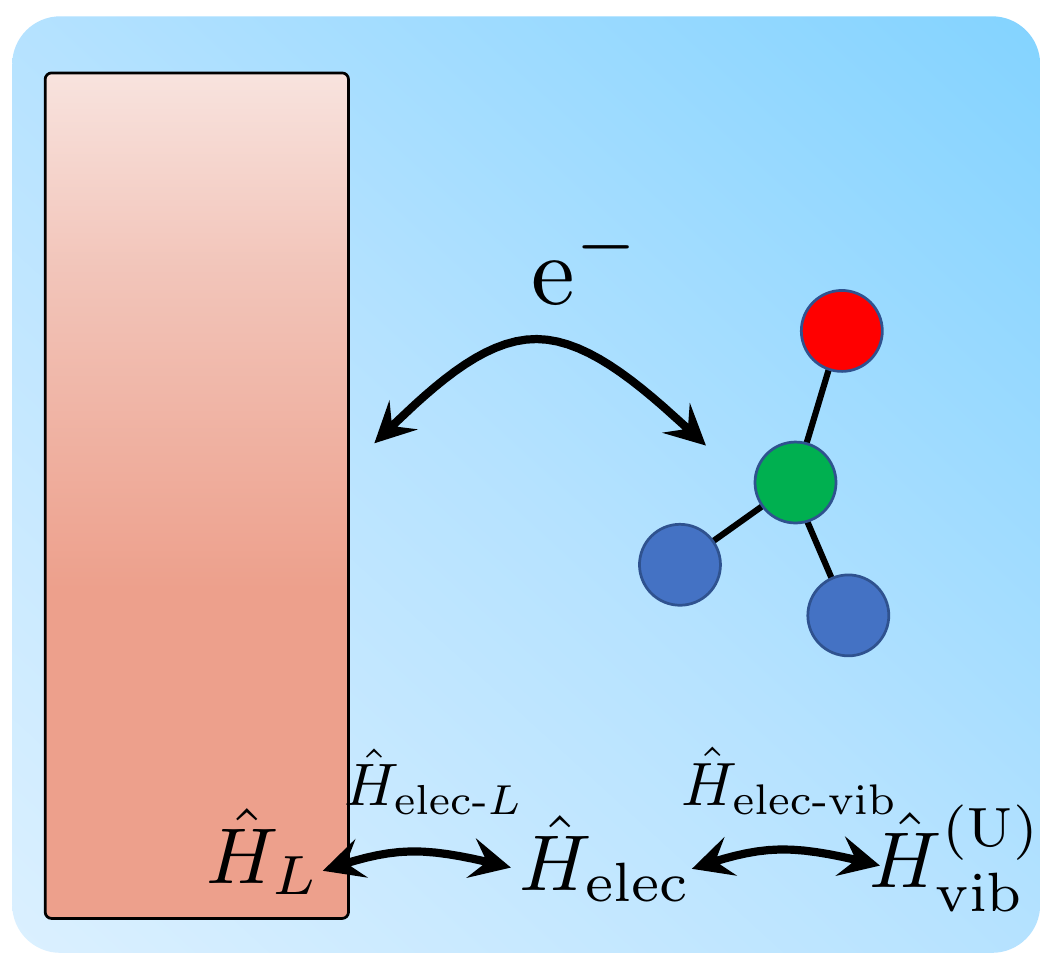}
    \end{center}
    \vspace{-5mm}
    \caption{Schematic representation of the type of Anderson-Holstein models studied in this work. The lead is described by $\hat{H}_L$, the initially uncharged molecular system by $\hat{H}_M = \hat{H}_{\textrm{elec}} + \hat{H}^{(U)}_{\textrm{vib}} + \hat{H}_{\textrm{elec-vib}}$, and the coupling between the two by $\hat{H}_{\textrm{elec-}L}$.} 
    \label{fig:ModelHam}
\end{figure}

We focus on the dynamics arising from a Hamiltonian of the Anderson-Holstein form,
\begin{equation}
\label{eq:total_ham}
\hat{H} = \underbrace{\hat{H}_{\textrm{elec}} + \hat{H}^{(U)}_{\textrm{vib}} + \hat{H}_{\textrm{elec-vib}}}_{\hat{H}_M} + \hat{H}_L + \hat{H}_{\textrm{elec-}L}.
\end{equation}
In particular, here we consider the case of a molecule $\hat{H}_M$ coupled to a single lead (electrode) with Hamiltonian $\hat{H}_L$ by $\hat{H}_{\textrm{elec-}L}$. The molecular Hamiltonian is comprised of the electronic states, vibrational states, and the coupling between them described by $\hat{H}_{\textrm{elec}}$, $\hat{H}^{(U)}_{\textrm{vib}}$, and $\hat{H}_{\textrm{elec-vib}}$ respectively. The molecule receives charge through its interaction with the lead which can alter the charge from its initial ``uncharged" state to a charged state. This system is shown schematically in Fig.~\ref{fig:ModelHam}. 

The particular form of the Anderson-Holstein Hamiltonian in Eq.~\ref{eq:total_ham} (also sometimes referred to as the Anderson-Newns model\cite{NEWNS1969}) has been previously been used to describe electron and proton transfer in electrochemical reactions\cite{Navrotskaya2009} as well as adsorption/desorption from metal interfaces\cite{Erpenbeck2018}. In the most general form, the molecular Hamiltonian can be written as a sum of the charged (C) and uncharged (U) states as
\begin{eqnarray}
\hat{H}_M &=& H^{(C)}_{\textrm{vib}}(\hat{\boldsymbol{Q}},\hat{\boldsymbol{P}}) \hat{d}^{\dagger}\hat{d} + H^{(U)}_{\textrm{vib}}(\hat{\boldsymbol{Q}},\hat{\boldsymbol{P}})[1-\hat{d}^{\dagger}\hat{d}],
\label{eq:Mol_Ham}
\end{eqnarray}
where
\begin{equation}
H^{(U)}_{\textrm{vib}}(\hat{\boldsymbol{Q}},\hat{\boldsymbol{P}}) = \frac{1}{2}\hat{\boldsymbol{P}}^T\hat{\boldsymbol{M}}^{-1}\hat{\boldsymbol{P}} + V^{(U)}(\hat{\boldsymbol{Q}}),
\end{equation}
and
\begin{equation}
H^{(C)}_{\textrm{vib}} (\hat{\boldsymbol{Q}},\hat{\boldsymbol{P}}) = \frac{1}{2}\hat{\boldsymbol{P}}^T\hat{\boldsymbol{M}}^{-1}\hat{\boldsymbol{P}} + V^{(C)}(\hat{\boldsymbol{Q}}) +\epsilon_M.
\end{equation}
The set of operators $  \boldsymbol{ \hat{Q}} $ and $ \boldsymbol{\hat{P}} $ describe the position and momentum of the nuclear degrees of freedom and $\epsilon_M$ is the energy difference between the charged and uncharged states. The diagonal mass matrix is denoted by $\boldsymbol{M}$ and the fermionic raising and lowering operators corresponding to the molecule are denoted by $\hat{d}$ and $\hat{d}^{\dagger}$ which satisfy the anticommutation rule $[\hat{d},\hat{d}^{\dagger}]_{+}=1$. Eq.~(\ref{eq:Mol_Ham}) can be rewritten as
\begin{equation}
\hat{H}_M = \epsilon_M d^{\dagger}d + H^{(U)}_{\textrm{vib}}(\hat{\boldsymbol{Q}},\hat{\boldsymbol{P}}) + \Delta V(\hat{\boldsymbol{Q}}) d^{\dagger}d,
\end{equation}
where we have defined $\Delta V(\boldsymbol{\hat{Q}})=V^{(C)}(\hat{\boldsymbol{Q}}) - V^{(U)}(\hat{\boldsymbol{Q}})$. By rewriting Eq.~(\ref{eq:Mol_Ham}) in this form and comparing with Eq.~(\ref{eq:total_ham}) one can see that
\begin{equation}
\hat{H}_{\textrm{elec}} = \epsilon_M  \hat{d}^{\dagger}\hat{d},
\end{equation}
and
\begin{equation}
\hat{H}_{\textrm{elec-vib}} = \Delta V(\hat{\boldsymbol{Q}}) \hat{d}^{\dagger}\hat{d}.
\end{equation}
The forms of $V^{(U)}(\hat{\boldsymbol{Q}})$ and $V^{(C)}(\hat{\boldsymbol{Q}})$ are kept general for now but will be given specified forms in Sec.~\ref{sec:Init}. Although in the present work we will use a simple form of the molecular Hamiltonian to allow for comparison to exact results we emphasize that the mapping approach to treating the fermionic dynamics described in Sec.~\ref{sec:mappingformalism} is fully compatible with atomistic treatments of molecular vibrations using diabatization schemes.\cite{Cave1996,Senthilkumar2003,Kondov2007,Difley2011,Oberhofer2012,Mao2019,Mao2020}

Here the lead is assumed to be comprised of free electrons giving the specific form of the Hamiltonian
\begin{equation}
\Hat{H}_L = \sum^G_k \epsilon_k \hat{c}^{\dagger}_k\hat{c}_k,
\end{equation}
and the lead-molecule coupling is taken to be via an exchange mechanism of the form
\begin{equation}
\label{eq:Ham_coupling}
\hat{H}_C = \sum^G_k t_k(\hat{d}^{\dagger}\hat{c}_k + \hat{c}^{\dagger}_k\hat{d}).
\end{equation}
The annihilation and creation operators $\hat{c}_k$ and $\hat{c}^{\dagger}_k$ destroy and create electrons of energy $\epsilon_k$ in the lead and satisfy $[\hat{c}_k,\hat{c}_{k'}^{\dagger}]_{+}=\delta_{kk'}$ while $\{t_k\}$ are the transfer coefficients between the lead states and the electronic states of the molecule. The distribution of fermionic states in the lead is determined by its spectral function $J_L(\epsilon)$ which allows the flexibility to treat leads ranging from metals to semiconductors.

\subsection{Classical mapping of a many-body bosonic and fermionic system}
\label{sec:mappingformalism}

While it has been long established how to classically describe bosonic degrees of freedom via the formal relationship between ladder operators and Cartesian variables\cite{Meyer1980,Stock1997} the case for fermions has faced considerably more difficulties. Previous work in this area has made use of quaternions to map fermionic operators into classical variables.\cite{Li2012,Li2013,Li2014,Levy2019} Recently it was demonstrated that for non-interacting fermionic systems it is possible to exactly replace the fermionic operators by classical-like phase space variables using the MM mapping\cite{Sun2021,MontoyaCastillo2023} giving the relations
\begin{equation}
\label{eq:c}
\hat{c} \to \frac{1}{\sqrt{2}}(q+ip),
\end{equation}
\begin{equation}
\label{eq:cdag}
\hat{c}^{\dagger} \to \frac{1}{\sqrt{2}}(q-ip).
\end{equation}
This mapping and the resulting classical equations of motion for $q$ and $p$ exactly reproduce the evolution of the one-body density matrix for noninteracting fermionic systems while outside of this limit, it is an approximation. It is important to emphasize that the classical oscillators of the MM mapping are used here to count the occupation of the single-particle fermionic states whose sum adds up to the total number of particles which differs from its traditional use in counting the total electronic population which always sums to one. Applying this mapping to the Hamiltonian of the Anderson-Holstein form in Eq.~(\ref{eq:total_ham}) gives
\begin{align}
\label{eq:mapped}
H(\boldsymbol{q},\boldsymbol{p},\boldsymbol{Q},\boldsymbol{P}) = \frac{1}{2}U(\boldsymbol{Q})(q^2_d + p^2_d-2\gamma_M)  + H^{(U)}_{\textrm{vib}}(\boldsymbol{Q},\boldsymbol{P})\nonumber \\ +  \sum^G_k \frac{\epsilon_k}{2}(q^2_k + p^2_k-2\gamma_L)  +  \sum^G_k t_k (q_dq_k+p_dp_k),
\end{align}
with
\begin{equation}
U(\boldsymbol{Q}) = \Delta V(\boldsymbol{Q}) + \epsilon_M.
\end{equation}
Here $\boldsymbol{q},\boldsymbol{p}$ are the mapped Cartesian coordinates that describe the fermionic variables of both the molecule ($q_{d}, p_{d}$) and lead ($\{q_{k}, p_{k}\}$). In Eq.~(\ref{eq:mapped}) we have also replaced the nuclear coordinates with classical variables $\boldsymbol{Q},\boldsymbol{P}$ which is a separate approximation. 

Each of these degrees of freedom evolves classically according to Hamilton's equations of motion
\begin{equation}
\label{eq:qd_dot}
\dot{q}_d = \frac{dH}{dp_d} = Up_d + \sum^G_kt_kp_k, 
\end{equation}
\begin{equation}
\dot{p}_d = -\frac{dH}{dq_d} = -Uq_d - \sum^G_kt_kq_k,
\end{equation}
\begin{equation}
\label{eq:Q0_dot}
\dot{Q}_j = \frac{dH}{dP_j} =   \frac{dH^{(U)}_{\textrm{vib}}}{dP_j},
\end{equation}
\begin{align}
\label{eq:P0_dot}
\dot{P}_j = -\frac{dH}{dQ_j} = -\frac{1}{2}\frac{dU}{dQ_j}(q^2_d + p^2_d-2\gamma_M) -  \frac{dH^{(U)}_{\textrm{vib}}}{dQ_j} \nonumber \\ - \sum^G_k \frac{dt_k}{dQ_j}(q_dq_k+p_dp_k),
\end{align}
\begin{equation}
\dot{q}_k = \frac{dH}{dp_k} = \epsilon_k p_k + t_kp_d,
\end{equation}
\begin{equation}
\label{eq:pk_dot}
 \dot{p}_k = -\frac{dH}{dq_k} = -\epsilon_k q_k - t_kq_d,
\end{equation}
where the dot refers to the derivative with respect to time. The second line of Eq.~(\ref{eq:P0_dot}) accounts for the fact that the transfer coefficients may depend on the nuclear coordinates, which can easily be incorporated within the mapping framework but is not studied in this work. 

The factors $\gamma_M$ and $\gamma_L$ parameterize the zero-point energy of the molecular and lead fermionic states respectively. Only for the choice of $\gamma_M=\gamma_L = \gamma$ does the MM mapping exactly describe the Anderson-Holstein model in the absence of nuclear modes, thus we will limit our attention to this case. The value of $\gamma$ has traditionally been restricted to lie between $0$ and $1$,\cite{Stock1997,Golosov2001,Cotton2013,Sun2021} although recent work has discussed the use of negative values.\cite{He2021} For noninteracting systems the classical equations of motion are independent of $\gamma$ and it only affects the initial conditions (see Sec.~\ref{sec:Init}). However, for interacting systems such as the case studied here the choice of this $\gamma$ affects the resulting dynamics as can be seen in Eq.~(\ref{eq:P0_dot}). For the particular value of $\gamma$, we investigated values of $\gamma=(\sqrt{3}-1)/2$\cite{Cotton2013,Runeson2019}, which gives the best result for 2-state systems such as the spin-boson problem, and $\gamma=0$, which has been shown to be the best choice as the number of states becomes large.\cite{Runeson2019} Overall we find that the latter of these choices yields slightly improved results over a wider range of Hamiltonian parameters and thus are shown in Sec.~\ref{sec:NumRes} while the results using the former are provided in SI Sec.~I. 

Upon integrating the equations of motion given in Eqs.~(\ref{eq:qd_dot})-(\ref{eq:pk_dot}) one can use the trajectories to compute time-dependent quantities such as the electronic population of the molecule
\begin{equation}
\label{eq:pop}
n_M(t) = \langle\hat{d}^{\dagger}\hat{d}\rangle \to  \frac{1}{2}\langle q^2_d + p^2_d -2\gamma \rangle,
\end{equation}
and the current
\begin{equation}
\label{eq:Current}
I(t) = -\frac{d}{dt}\sum_k\langle \hat{c}^{\dagger}_k\hat{c}_k \rangle \to -\sum_k t_k\langle p_dq_k - q_dp_k \rangle.
\end{equation}
where the mapping from the fermionic operators to the classical variables are obtained by using Eqs.~(\ref{eq:c}) and (\ref{eq:cdag}). Here the angular brackets denote averaging with respect to the initial density matrix. Appendix A shows that the MM mapping guarantees the total charge and energy are conserved just as is the case with the exact quantum dynamics.

As discussed in the introduction, the MM mapping is not the only choice of a classical mapping that is available for fermionic operators. Indeed, the LMM\cite{Li2012} and its generalized form known as the complete quasiclassical map (CQM)\cite{Levy2019} are also exact in the same limits as the MM mapping. It was also shown in Ref.~\citenum{Sun2021} that each of these mappings perform similarly outside of the non-interacting fermion case. The reason for this is likely due to the fact that despite each of these mappings trying to capture fermionic properties they can actually be expressed in a unified mapping framework, along with the MM mapping, which makes no assumptions of fermionic canonical commutation relations \cite{Liu2016,Liu2017}. In SI Sec.~III we compute the results obtained using the CQM approach for the Anderson-Holstein model we investigate here and demonstrate that it gives almost numerically indistinguishable results to those obtained using MM mapping. Since MM mapping only requires half the number of degrees of freedom to describe the same system as the LMM and CQM in Sec.~\ref{sec:NumRes}, we show the results obtained using MM mapping.

\section{Model system and generation of initial conditions}
\label{sec:Init}

To examine the ability of the MM mapping to describe the many-body physics of the Anderson-Holstein model we specify the nuclear vibrational Hamiltonians in Eq.~(\ref{eq:total_ham}) to take the forms,
\begin{equation}
H^{(U)}_{\textrm{vib}} = \frac{1}{2}\hbar\omega(P^2 + Q^2),
\end{equation}
and
\begin{eqnarray}
\label{eq:H_charged}
H^{(C)}_{\textrm{vib}} &=& \frac{1}{2}\hbar\omega(P^2 + Q^2) + \sqrt{2}gQ + \epsilon_M, \nonumber \\ &=&\frac{1}{2}\hbar\omega \left[P^2 + \left(Q+\frac{\sqrt{2}g}{\hbar\omega}\right)^2\right] + \tilde{\epsilon}_M.
\end{eqnarray}
In our case, the nuclear Hamiltonian thus consists of one harmonic vibrational degree of freedom with frequency $\omega$ coupled to the two electronic states of the molecule where $g$ mediates the electron-vibration coupling, $\epsilon_M$ is the bias of the charged state and $\tilde{\epsilon}_M = \epsilon_M - g^2/\hbar\omega$ is the renormalized molecular energy in the presence of the vibrational mode. This model was previously investigated using the CME-SH approach that we compare to here.\cite{Dou2015} The initial density matrix is
\begin{equation}
\rho = \rho_M \otimes \rho^{\textrm{eq}}_{\textrm{vib}}  \otimes \rho^{\textrm{eq}}_L, 
\end{equation}
where $\rho_M$ represents the initial density of the fermionic degrees of freedom of the molecule where the molecule is initialized such that only $H^{(U)}_{\textrm{vib}}$ is populated i.e. $n_M(0)=0$ (see Eq.~\ref{eq:pop}). The nuclear vibrational degrees of freedom are thermalized in this state according to the Wigner distribution given by,
\begin{equation}
\rho^{\textrm{eq}}_{\textrm{vib}} = \frac{\alpha}{\pi} e^{-\alpha(Q^2 + P^2)},
\end{equation}
where $\alpha = \tanh(\beta\hbar\omega/2)$ and $\beta=1/( k_{B}T)$ is the inverse temperature. In the limit of $\beta\hbar\to 0$ this reduces to the classical Boltzmann distribution.

$\rho^{\textrm{eq}}_L$ is the thermal equilibrium density matrix of
the lead states. Since the electrons in the lead are in thermal equilibrium, the population of a state with energy $\epsilon$ is given by the Fermi-Dirac distribution
\begin{equation}
n_f(\epsilon) = \frac{1}{1+e^{\beta(\epsilon-\mu)}},
\end{equation}
where $\mu$ is the chemical potential. Here, we assume a lead that is metallic in nature such that the fermionic states of the lead form a continuum and are described by the continuous spectral function
\begin{equation}
J_L(\epsilon) = \frac{\Gamma}{(1+e^{A(\epsilon-B/2)})(1+e^{-A(\epsilon+B/2)})},
\end{equation}
where $A$ and $B$ define the cutoff and width of the lead distribution and $\Gamma$ controls the coupling strength of the lead to the molecule. The transfer coefficients, $t_k$ (see Eq.~\ref{eq:Ham_coupling}), are modeled in the wide band limit following the procedure of Refs.~\citenum{Li2014} and \citenum{Levy2019} giving
\begin{equation}
t_k = \sqrt{\frac{J_L(\epsilon_k)\Delta \epsilon }{2\pi}},
\end{equation}
where $\Delta \epsilon$ is the spacing between lead states. The initial conditions for the molecular fermionic degrees of freedom are
\begin{equation}
q_d(0) = \sqrt{2\gamma + 2n_M(0)}\cos(\theta),
\end{equation}
and
\begin{equation}
p_d(0) = -\sqrt{2\gamma + 2n_M(0)}\sin(\theta) ,
\end{equation}
where $\theta$ is a uniform random number between $0$ and $2\pi$ and $n_M(0)$ is initial charge state of the molecule. The lead degrees of freedom are initialized similarly as
\begin{equation}
q_k(0) = \sqrt{2\gamma + 2n_k}\cos(\theta), 
\end{equation}
and
\begin{equation}
p_k(0) = - \sqrt{2\gamma + 2n_k}\sin(\theta), 
\end{equation}
where the $n_k$ are sampled using the procedure of Ref.~\onlinecite{Levy2019} where each $n_k$ is either zero or one such that the Fermi-Dirac distribution is reproduced upon ensemble averaging. In what follows all units are reported in atomic units unless otherwise stated. Additional Computational details can be found in the SI Sec.~IV.

\section{Results}
\label{sec:NumRes}

To investigate the performance of the MM mapping in describing the Anderson-Holstein model we first look at cases without the presence of the vibrational mode, in which the mapping should yield the exact result regardless of the strength of the lead-molecule coupling\cite{Sun2021,MontoyaCastillo2023}. We then consider the effect of coupling this system to a vibrational mode of the lead and how this alters the ability to capture the properties of the system.

\subsection{Lead-molecule system}
\label{sec:lead-molecule}
When $g=0$ in Eq.~(\ref{eq:H_charged}) the nuclear vibrational mode is decoupled from the system and it reduces to an Anderson model consisting of just the lead and molecule (often referred to as a dot). 
In the limit of small lead-molecule coupling, $\Gamma$, the equilibrium molecular population in the presence of a wide band lead is given by\cite{Bruus_Henrik2004,Stefanucci2013-ju}
\begin{equation}
\label{eq:eq_pop_analytic}
n^{eq}_M = \int \frac{d\epsilon}{2\pi} \frac{\Gamma}{(\epsilon-\tilde{\epsilon}_M)^2 + (\Gamma/2)^2} n_f(\epsilon).
\end{equation}
 In the case of $g\to0$ (no electron-vibration coupling) Eq.~(\ref{eq:eq_pop_analytic}) is exact for all values of $\Gamma$. Using Eq.~(\ref{eq:eq_pop_analytic}) we can test how well the classical dynamics of the MM mapping recovers equilibrium properties both in the absence and presence of nuclear motion as a function of the lead-molecule coupling strength. Since, in the absence of nuclear modes the MM mapping gives a quadratic Hamiltonian, which ensures the dynamics are exact, it should recover the correct equilibrium population regardless of the lead-molecule coupling strength. 
 
\begin{figure}[h]
    \begin{center}
        \includegraphics[width=0.5\textwidth]{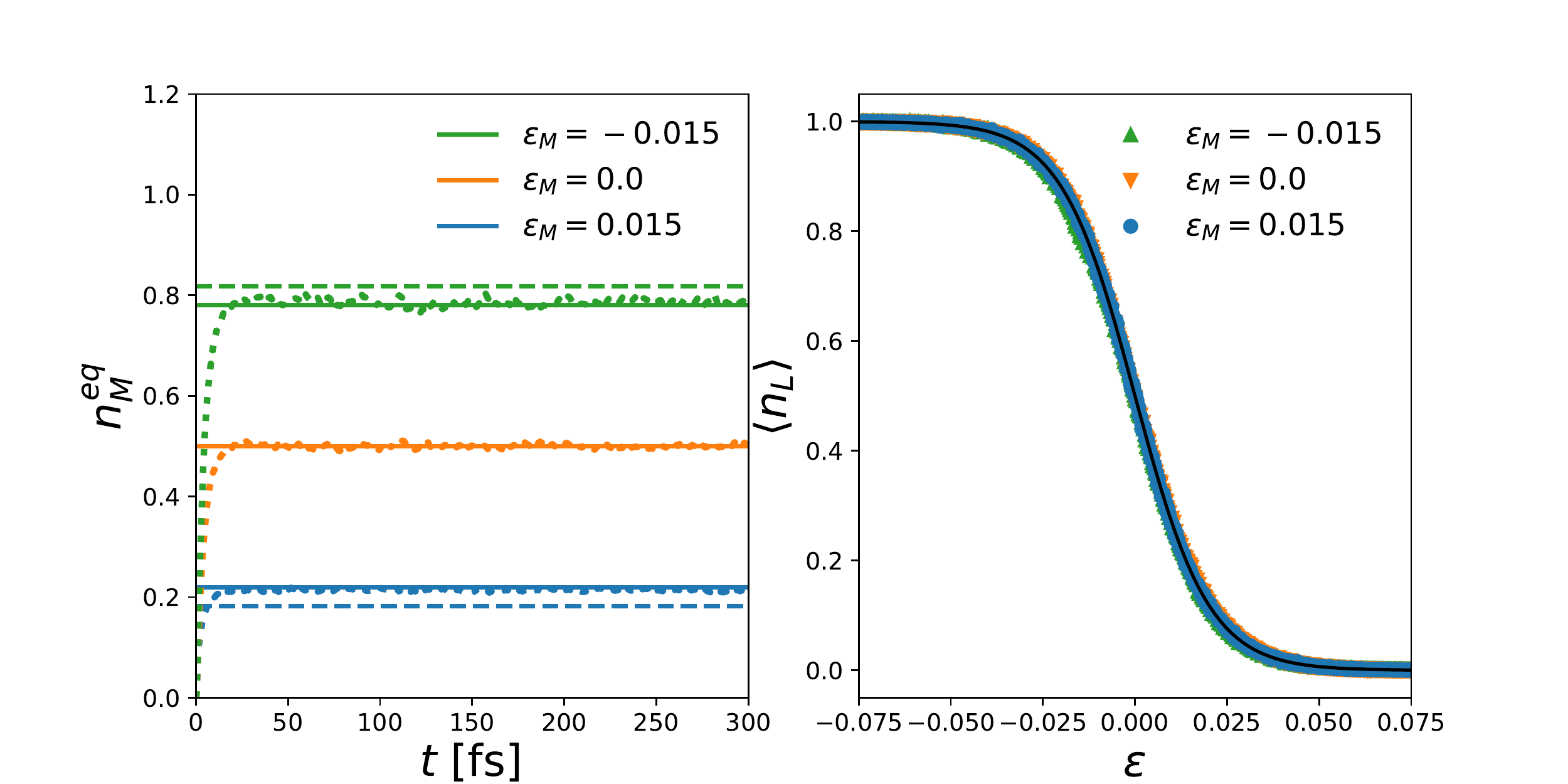}
    \end{center}
    \vspace{-5mm}
    \caption{Left: Transient dynamics of the molecule population with $\Gamma=0.006$, $\mu=0.0$ and $\beta=100$. Dotted lines are the transient populations computed using the MM mapping, dashed lines are the equilibrium value predicted by the CME-SH method without broadening and solid lines are the exact values obtained from Eq. (\ref{eq:eq_pop_analytic}). Right: Lead distribution after 750 fs. Colored shapes are distributions computed from the MM mapping and the solid black line is the Fermi-Dirac distribution. 10000 trajectories were used to obtain these results.} 
    \label{fig:eq_check}
\end{figure}

Figure~\ref{fig:eq_check} shows that in this case ($g=0$) the MM mapping is able to capture the correct equilibrium molecular population as the energy bias $\epsilon_M$ is scanned revealing that the hybridization between the molecular level and the lead is being properly described. This is not the case for the CME-SH methodology,\cite{Dou2015,Dou2015iii} which only recovers correct equilibrium populations after being ``broadened" to account for the hybridization between the lead and molecule. While the broadening schemes that have been used in conjunction with CME-SH are effective in correcting the equilibrium behavior in the absence of nuclear motion it is unclear how to extend them beyond cases where analytic results provide guidance. In contrast, the MM mapping naturally incorporates the hybridization through the explicit dynamics of the lead degrees of freedom and does not need to be modified to capture the equilibrium behavior in the absence of nuclear motions. In addition, the current (Eq.~(\ref{eq:Current})) is correctly zero when the system is in equilibrium.

\subsection{Lead-molecule-vibration system}

We now consider the case where the lead-molecule system is now coupled to a nuclear vibrational mode (see Fig.~\ref{fig:ModelHam}) which can dissipate energy. The strength of the coupling of the molecule's electronic population to the vibrational is controlled by $g$ (Eq.~\ref{eq:H_charged}). Due to the presence of the nuclear motion coupled to the molecular electronic degrees of freedom the MM representation is no longer exact. 

\begin{figure}[h!]
    \begin{center}
\includegraphics[width=0.5\textwidth]{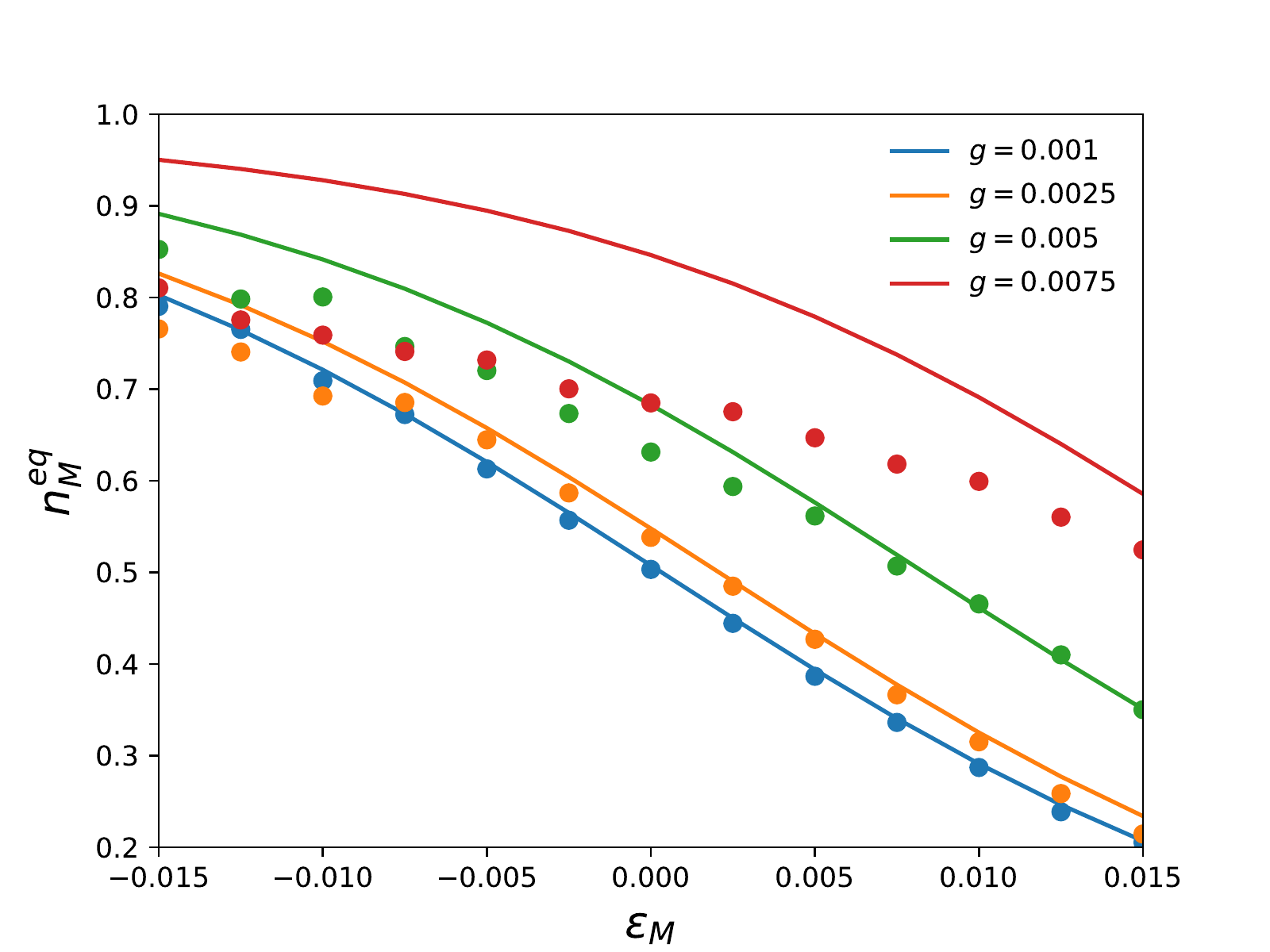}
    \end{center}
    \vspace{-5mm}
    \caption{Equilibrium molecule population as a function of $\epsilon_M$. $\Gamma=0.003$, $\mu=0.0$, $\hbar\omega=0.003$ and $\beta=100$. The solid line are the exact NRG results and dots are from the explicit MM mapping dynamics using 10000 trajectories. Here $\gamma=0$. $1200$ lead states were used ranging in energy from $-0.1$ to $0.1$ except for the case where $g=0.0075$ in which $2400$ states were used ranging from $-0.2$ to $0.2$.} 
    \label{fig:eq_eM_check}
\end{figure}

Figure~\ref{fig:eq_eM_check} shows the equilibrium molecule population (obtained at 1.5~ps) as a function of the energy bias for a series of electron-vibration coupling ($g$) values. The parameter regimes shown correspond to those in Ref.~\citenum{Dou2015} such that comparisons can be made to the CME-SH approach. Despite the MM mapping not being exact we see that it is able to accurately predict the correct population for the cases when $g=0.001$, $g=0.0025$, and $g=0.005$. However, for the strongest coupling to the vibrational mode $g=0.0075$ we find that the MM mapping underestimates the long-time population. We also observe that MM mapped dynamics more accurately capture the equilibrium molecule population at large positive bias ($\epsilon_{M}>0$) than negative bias. This is likely due to the fact that at high positive bias the molecule's energy is shifted high into the Fermi distribution where the population of lead states is much smaller which effectively reduces the amount of coupling between the lead and the molecule.

\begin{figure}[h!]
    \begin{center}
\includegraphics[width=0.45\textwidth]{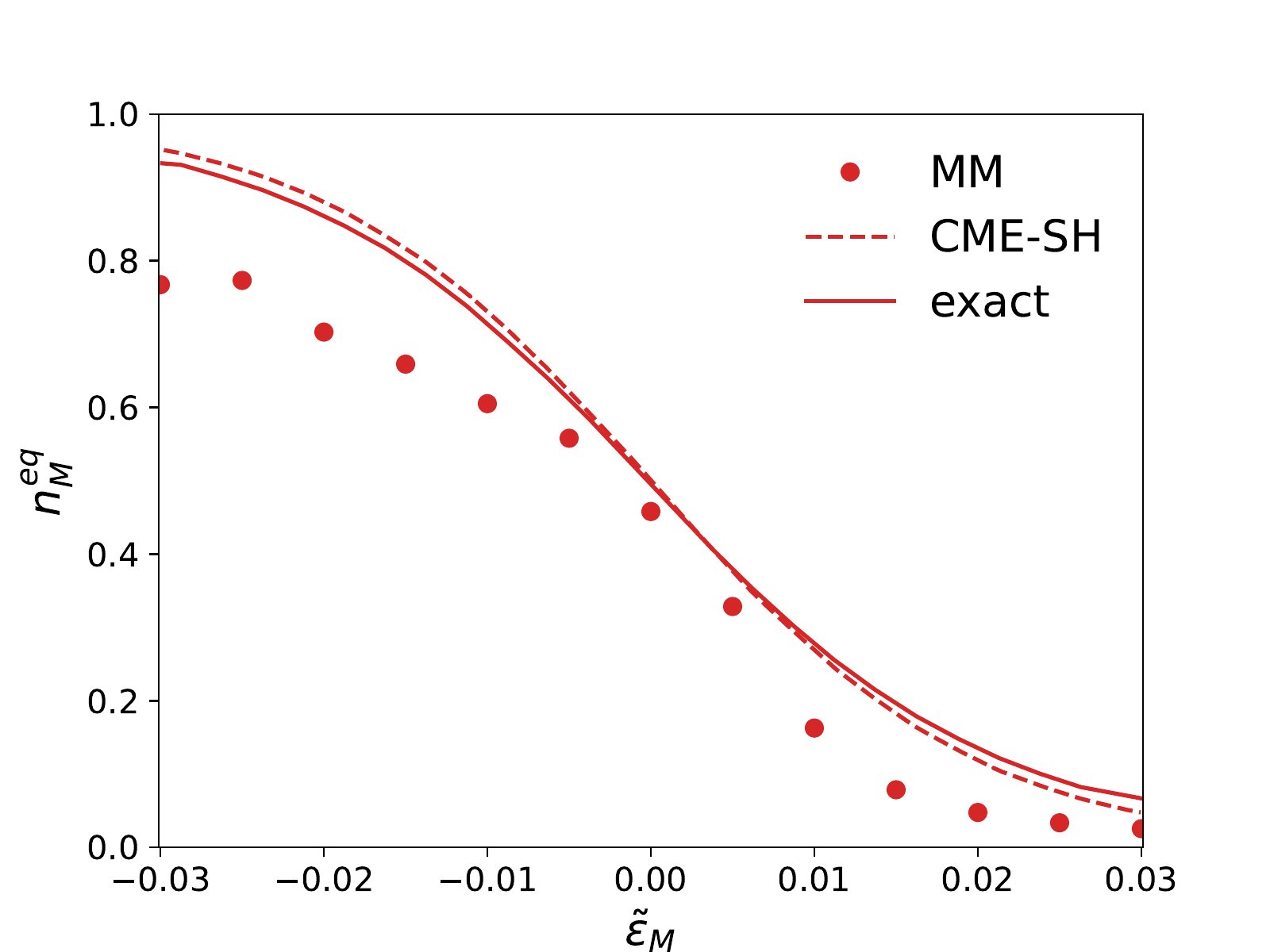}
\includegraphics[width=0.45\textwidth]{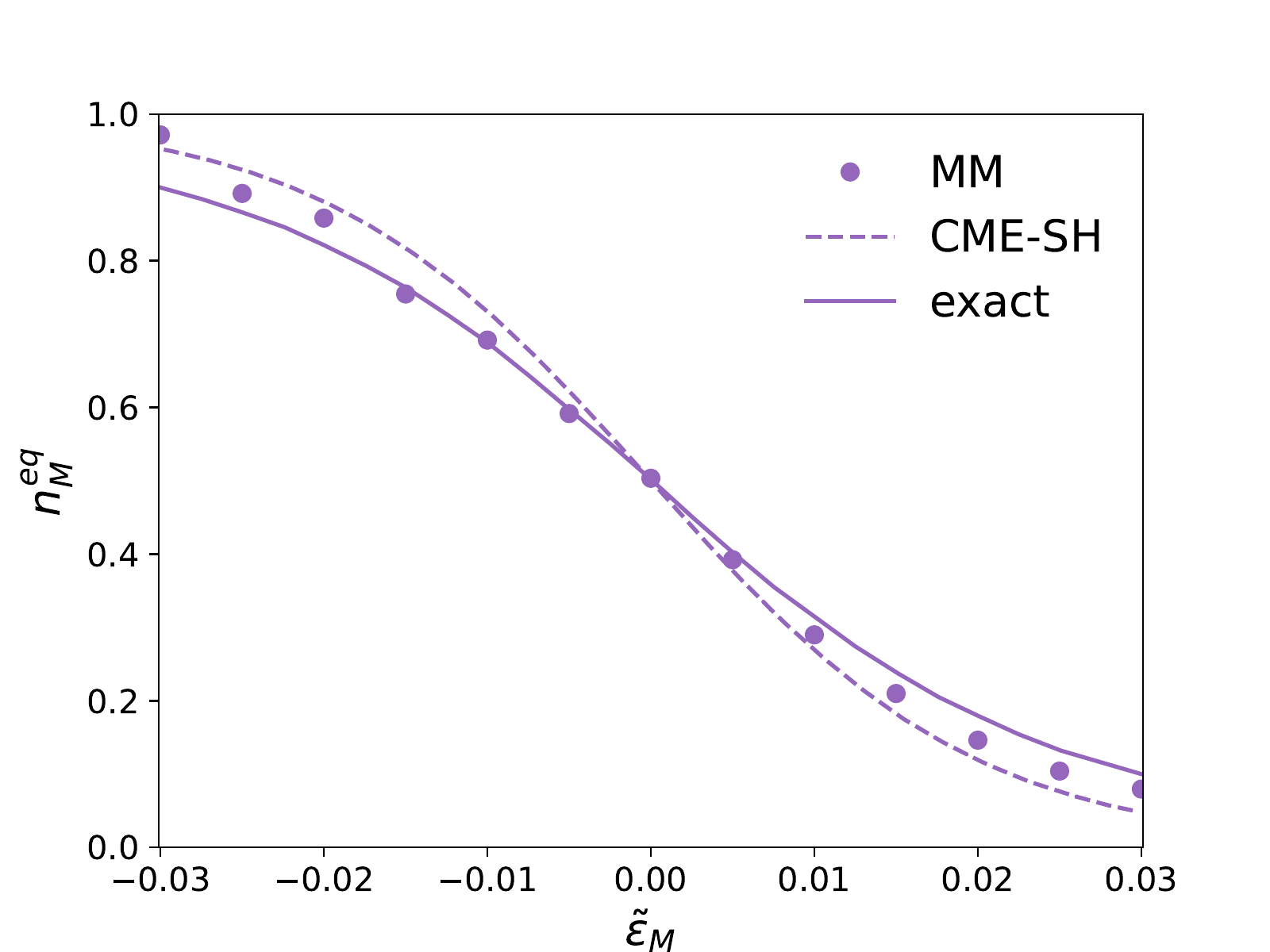}
    \end{center}
    \vspace{-5mm}
    \caption{Equilibrium molecule population as a function of $\tilde{\epsilon}_M$. $\mu=0.0$, $\hbar\omega=0.003$ and $\beta=100$. Top panel: $\Gamma=0.003$ and $g=0.075$ Bottom panel: $\Gamma=0.01$ and $g=0.025$  Solid lines are the NRG results, dots are from the MM mapping using 10000 trajectories and dashed lines are unbroadened CME-SH results taken from Ref. \citenum{Dou2015}. Broadened CME-SH results are indistinguishable from the NRG results and are not shown.} 
    \label{fig:eq_eM_check2}
\end{figure}

To further test MM mapping's ability to capture the correct long-time molecular population in Fig.~\ref{fig:eq_eM_check2} we plot the molecular population as a function of the renormalized bias, $\tilde{\epsilon}_M$, for a case with strong vibrational coupling and weak lead coupling (top panel) and a case with strong lead coupling and weak vibrational coupling (bottom panel). Our MM mapping results are compared with the numerically exact numerical renormalization group (NRG) and CME-SH calculations without broadening in Fig.~\ref{fig:eq_eM_check2}. As expected from the observations in Fig.~\ref{fig:eq_eM_check} the MM mapping is not able to reproduce the correct equilibrium molecule population in the case of strong vibrational coupling but is very accurate for the case of weak vibrational coupling even when the lead is strongly coupled to the molecule. The latter result follows from the fact that the MM mapping describes the bare fermionic system exactly regardless of the value of $\Gamma$ (see Sec.~\ref{sec:lead-molecule}). A weakly coupled nuclear mode does not significantly affect this description and demonstrates that there are regimes in which a mapping description that employs fully linearized trajectories may be more accurate than the CME-SH since there is no ambiguity in how to broaden the results. However, in cases with strong coupling to the vibrational mode, even without broadening, the CME-SH is able to more accurately capture the exact result. The failure of mean-field dynamics and related methods such as classical mappings to capture detailed balance for electronic systems coupled to bosonic modes has been well documented in the literature and again presents a challenge in the cases of stronger coupling to the vibrational mode.\cite{Bellonzi2016,Dou2016,Jain2017} 

\begin{figure}[h]
    \begin{center}
        \includegraphics[width=0.5\textwidth]{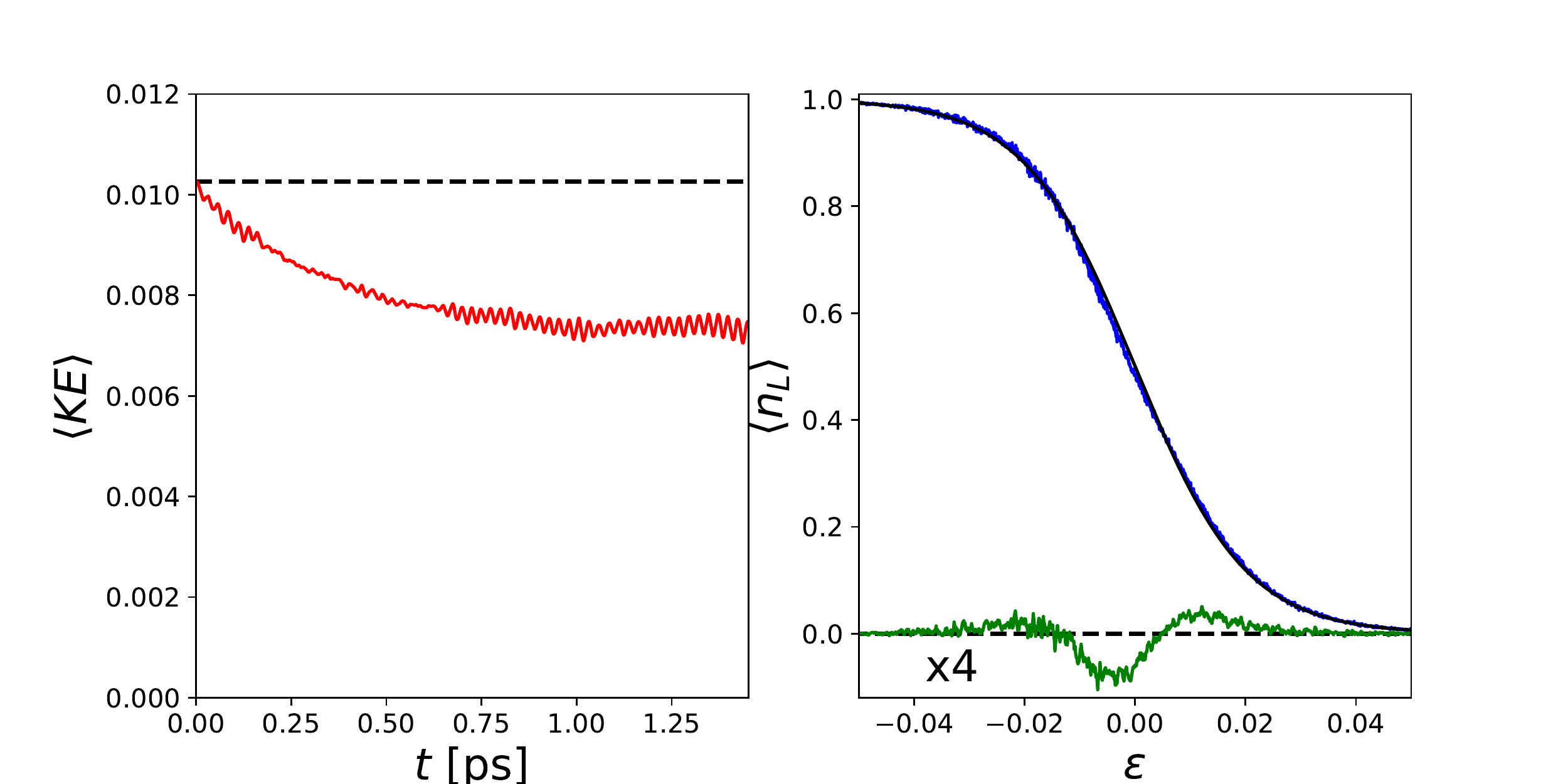}
    \end{center}
    \vspace{-5mm}
    \caption{The molecule with zero bias and the lead are both initialized in thermal equilibrium. Left: the average kinetic energy from the MM mapping dynamics (red line) and exact value (black dashed line). Right: The initial (black line) and final (blue line) lead distribution. The green is the difference between the two and has been magnified by a factor of 4 for clarity. $\Gamma=0.01$, $\mu=0.0$, $\hbar\omega=0.003$, $g=0.025$, $\beta=100$. } 
    \label{fig:DB_test}
\end{figure}

To investigate the failure of the MM mapping in capturing detailed balance, we performed a calculation in which the molecule with no bias between the charged and uncharged states ($\tilde{\epsilon}=0$) and the lead are initialized in thermal equilibrium. Due to the absence of a bias between the charged states and since everything is at the same temperature there should be no net exchange of energy between the lead and molecular degrees of freedom and hence the kinetic energy should not evolve with time. The left panel of Fig.~\ref{fig:DB_test} shows the time evolution of the kinetic energy of the vibrational mode where we see that the MM mapping shows a drift in the average kinetic energy until it reaches a steady state value of $\sim75\%$ of its initial value. The right panel of Fig.~\ref{fig:DB_test} shows the initial (black) and final (blue) populations of the lead states where, after magnifying the difference between them (green), we observe that those near the value of $\tilde{\epsilon}=0$ have lost population which has been redistributed to the surrounding states. From the green line we see that more of the population has been pushed into states of higher energy. This increase of energy in the lead is responsible for the decrease in the average kinetic energy of the molecule. Similar results are found when the number of states in the lead is increased verifying that this effect is not an artifact of improper lead discretization (see SI. Sec.~II).

\section{Conclusion}

We have presented a simple and practical method for simulating electrochemical systems using classical trajectories. By utilizing the MM mapping to describe fermionic degrees of freedom we have shown it is possible to capture the fundamental physics of lead-molecule and electron-vibration coupling in the Anderson-Holstein model. Most notable is the mapping approach's ability to correctly capture the lead-molecule hybridization. The mapping approach thus provides advantages over other methods that could be applied to an atomistic description of the molecular degrees of freedom such as the CME-SH approach\cite{Dou2015,Dou2015iii} since it naturally captures the correct electronic population of the bare fermionic system without the need for an {\it ad hoc} broadening procedure. However, while surface hopping methods accurately capture detailed balance,\cite{Dou2015,Jain2017,Schmidt2008} we have shown that, consistent with other MM mapped systems, when strong coupling to the nuclear motion is present detailed balance issues arise leading to incorrect molecular populations. Given that a number of strategies have been introduced to improve the detailed balance failures of mapping approaches to semiclassical dynamics these provide a potential avenue for improvements in future work.

Ultimately we believe this straightforward procedure should be usable in a variety of contexts where strong molecule-lead and weak electron-vibration couplings are present making it complementary to other trajectory-based approaches as well as perturbative master equations. The flexibility that this method affords should make it possible to extend it to describe a range of electrochemically relevant properties such as adsorption, charge transfer in atomistic environments as well as bond breaking/formation at charged surfaces when combined with \emph{ab initio} descriptions of the forces.

\section*{Acknowledgments}
This work was supported by National Science Foundation Grant No. CHE-2154291.

\section*{Data Availability}
The data that supports the findings of this study are available within the article and its supplementary material.

\appendix

\section{Conservation of charge and energy}
\label{sec:chargeconservation}
The total charge is given by the sum of the fermionic population in the lead and the molecule
\begin{equation}
\hat{T} = \hat{d}^{\dagger}\hat{d} + \sum^G_k \hat{c}^{\dagger}_k\hat{c}_k.
\end{equation}
Using the Heisenberg equation of motion and Eq.~(\ref{eq:total_ham}) it is straightforward to show that
\begin{equation}
\frac{d}{dt}\hat{T} = i[\hat{H},\hat{T}] = 0,
\end{equation}
meaning that charge is conserved in this system when treated quantum mechanically. Likewise, it can also be shown that when the system is mapped classically using the MM mapping that charge is also conserved. Within the mapping the total charge is given by
\begin{equation}
T = \frac{1}{2}(q^2_d + p^2_d - 2\gamma) + \frac{1}{2}\sum^G_k(q^2_k + p^2_k - 2\gamma).
\end{equation}
From this definition and using the classical equations of motion given in Eqs. (\ref{eq:qd_dot})-(\ref{eq:pk_dot}) it follows that
\begin{equation}
\frac{d}{dt}T = q_d\dot{q}_d + p_d\dot{p}_d + \sum^G_k (q_k\dot{q}_k + p_k\dot{p}_k)=0.
\end{equation}
Similarly the total energy is conserved, which trivially follows from the fact that quantum Hamiltonian commutes with itself and Hamilton's equations of motion (Eqs.~(\ref{eq:qd_dot})-(\ref{eq:pk_dot})) conserve the total energy of the system.
These results reveal that regardless of the value of $\gamma$ the charge and energy of the total system is conserved on the trajectory level. While it is true that charge and energy are conserved at the trajectory level this does not imply that individual trajectories are physically meaningful. One must still average a sufficient number of configurations to properly approximate the quantum dynamics of the system.

\bibliography{EC}
\end{document}


\title{Supporting information for: Electron transfer at electrode interfaces via a straightforward quasiclassical fermionic mapping approach}

\author{Kenneth A. Jung}
\author{Joseph Kelly}
\author{Thomas E. Markland}
\email{tmarkland@stanford.edu}
\affiliation{Department of Chemistry, Stanford University, Stanford, California, 94305, USA}

\date{\today}

\maketitle

\tableofcontents

\newpage

\section{Alternative choice of the point energy}
\begin{figure}[h!]
    \begin{center}
\includegraphics[width=0.85\textwidth]{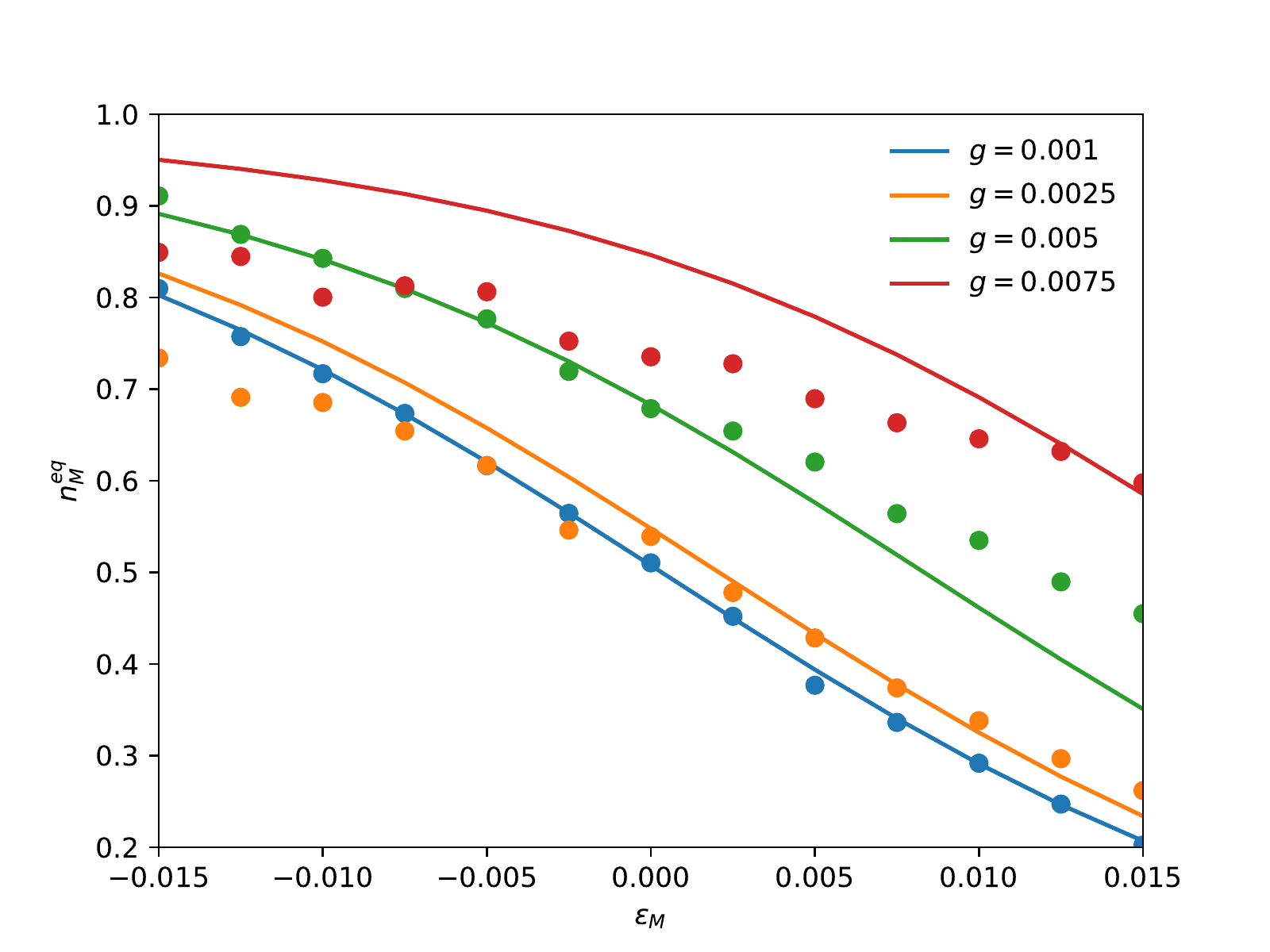}
    \end{center}
    \vspace{-5mm}
    \caption{Equilibrium molecule population as a function of $\epsilon_M$. Parameters are the same as those in Fig.~3 in the main text except here $\gamma=\frac{\sqrt{3}-1}{2}$} 
    \label{fig:standard_gamma}
\end{figure}

In this section we present molecular equilibrium populations of the Anderson-Holstein model for the choice of $\gamma=\frac{\sqrt{3}-1}{2}$ in figure \ref{fig:standard_gamma}.

\section{Band Density Convergence}
\begin{figure}[h]
    \begin{center}
        \includegraphics[width=0.85\textwidth]{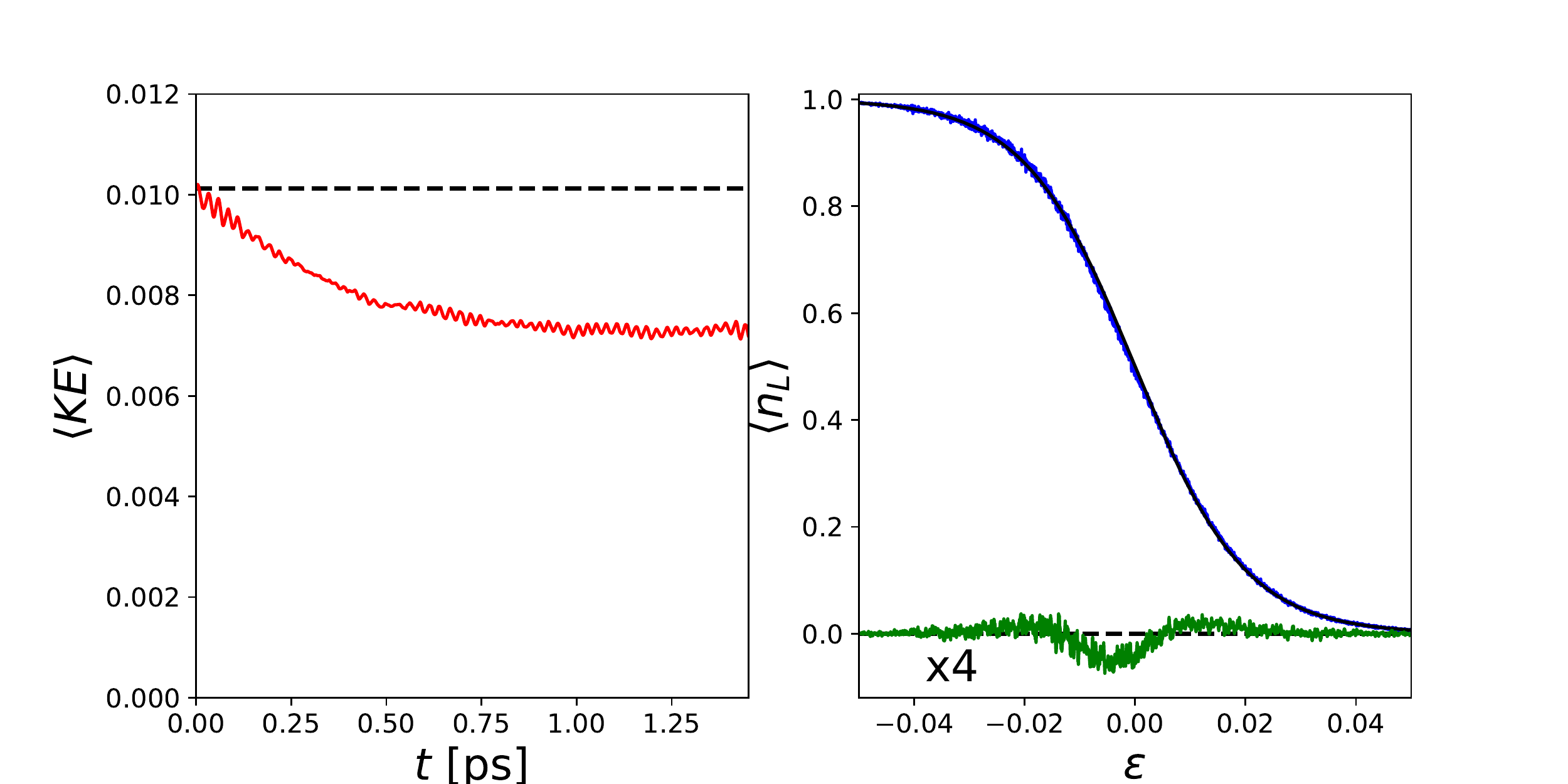}
    \end{center}
    \vspace{-5mm}
    \caption{The molecule with zero bias and the lead are both initialized in thermal equilibrium. Left: the average kinetic energy from the MM mapping dynamics (red line) and exact value (black dashed line). Right: The initial (black line) and final (blue line) lead distribution. The green is the difference between the two and has been magnified by a factor of 4 for clarity. $\Gamma=0.01$, $\mu=0.0$, $\hbar\omega=0.003$, $g=0.025$, $\beta=100$. 2400 lead states were used in this calculation compared with the 1200 states used in Fig.~5 in the main text.} 
    \label{fig:Band_test}
\end{figure}

To ensure our results are properly converged with respect to the number of electrode states Fig.~\ref{fig:Band_test} shows the same regime as in Fig.~5 of the paper but with double (2400) the number of states used to discretize the lead. From this figure, we see that although there are subtle differences in the transient dynamics of the average kinetic energy the overall value is still incorrectly time-dependent and decreases from its initial value by roughly $75\%$. Looking at the difference between the final and initial thermal lead occupations we see the same effect of the population being pushed into higher states. These results verify that the detailed balance issues observed are likely not artifacts of an improper discretization of the lead.

\section{Comparing the Meyer-Miller mapping and the complete quasiclassical mapping}
\begin{figure}[h]
    \begin{center}
        \includegraphics[width=0.85\textwidth]{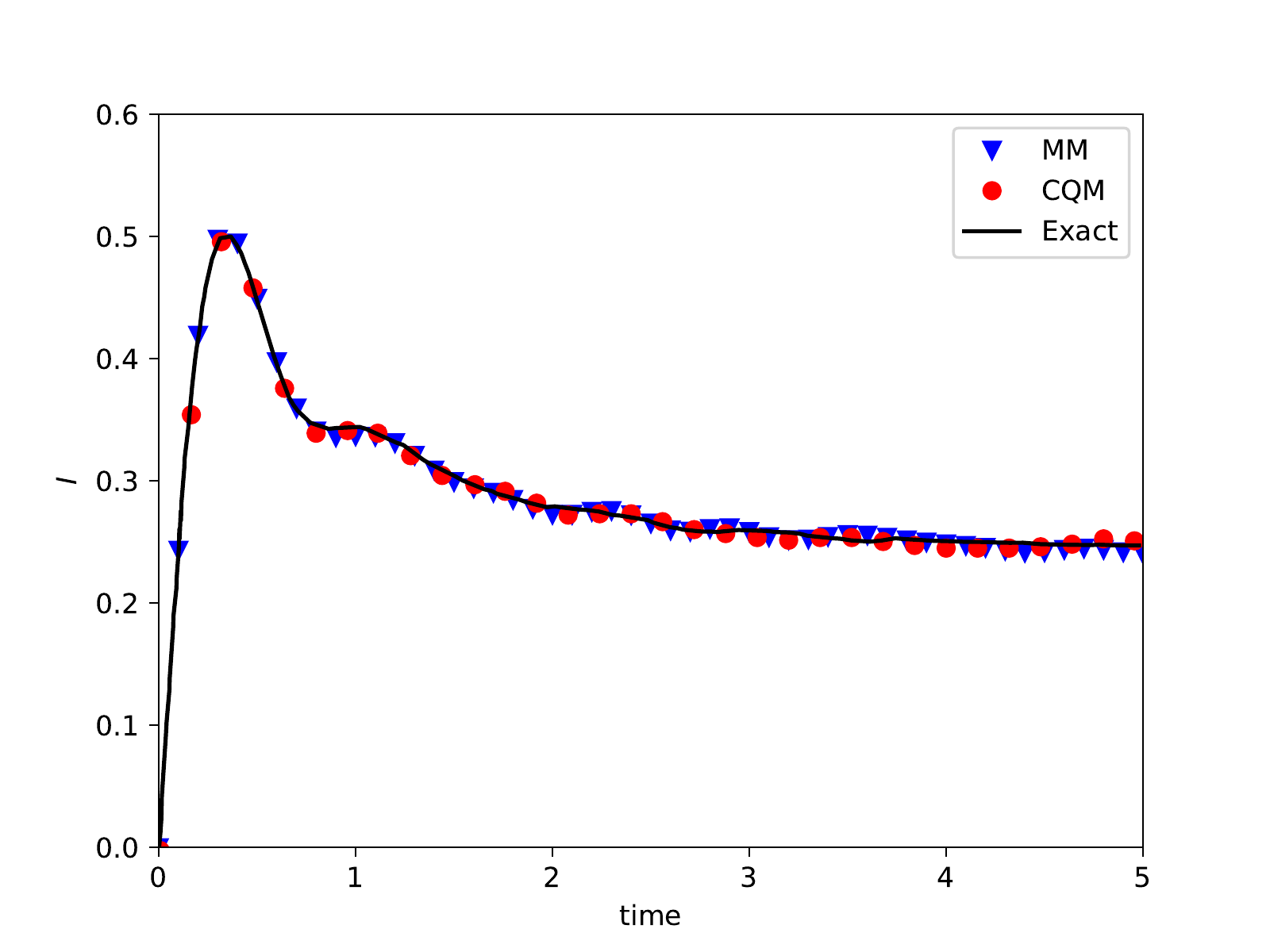}
    \end{center}
    \vspace{-5mm}
    \caption{Anderson model current as a function of time. The Hamiltonian and model parameters are taken from Ref.~\citenum{Levy2019}.} 
    \label{fig:CQM_compare}
\end{figure}

\begin{figure}[h]
    \begin{center}
        \includegraphics[width=0.85\textwidth]{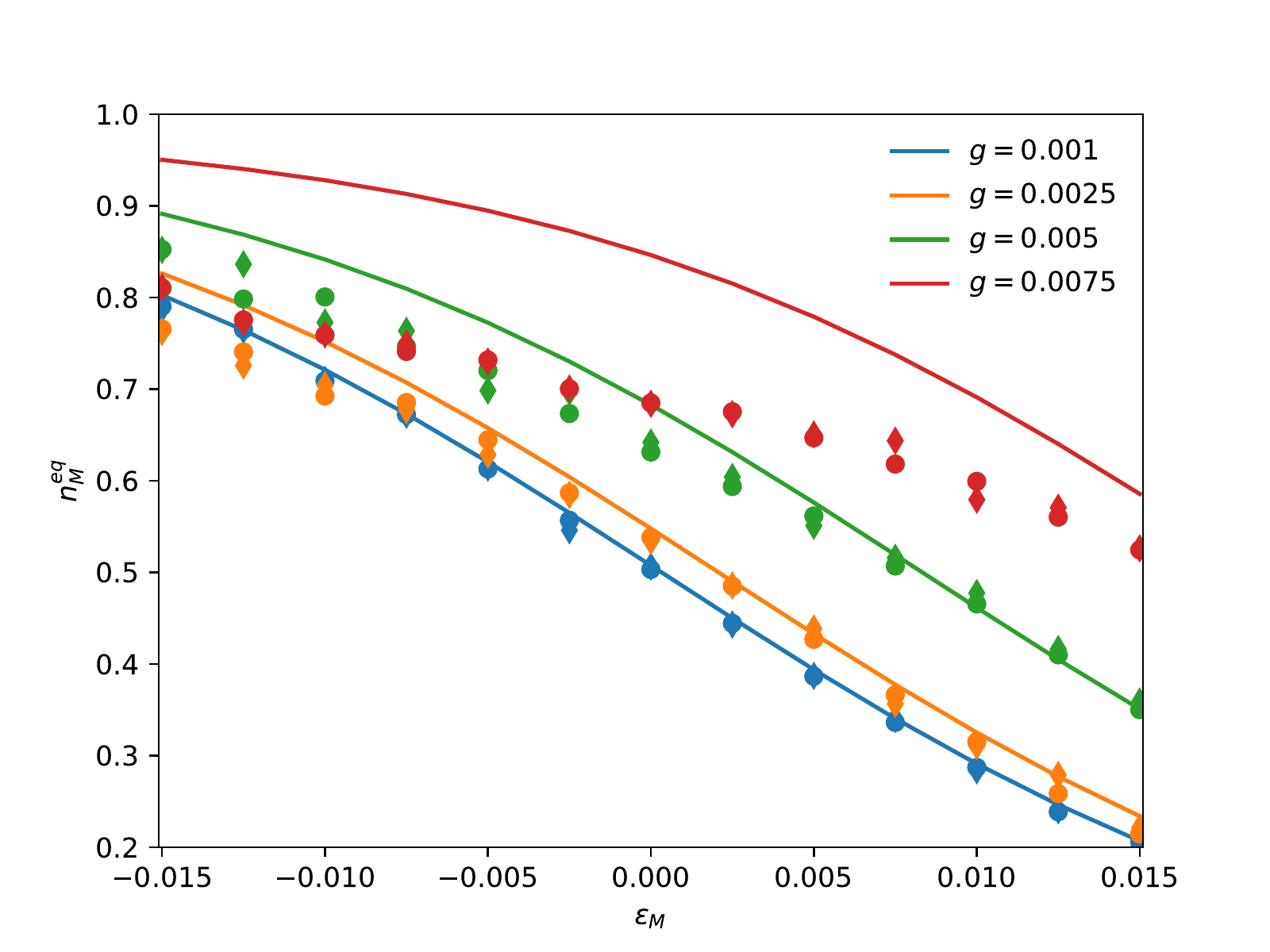}
    \end{center}
    \vspace{-5mm}
    \caption{Equilibrium molecule population of the MM mapping (circles) and CQM (diamonds) as a function of $\epsilon_M$. Parameters are the same as those in Fig.~3 in the main text.} 
    \label{fig:CQM_compare2}
\end{figure}

Here we report a comparison of the MM mapping and the CQM for two cases: the spinless Anderson transport model and the Anderson-Holstein model. As discussed in the main text both the MM mapping and the CQM are exact in the case of noninteracting electrons. We demonstrate this in Fig.~\ref{fig:CQM_compare} we see that, as expected, both mappings reproduce the time dependent current for the non-interacting model. For the case of the Anderson-Holstein model we see from Fig.~\ref{fig:CQM_compare2} that even for fermionic systems interacting with a vibrational mode the two methods are nearly identical. These results compliment those found when comparing the two mapping methods for the case of interacting fermions.\cite{Sun2021}

\section{Computational Details}
The equations of motion for all methods tested were integrated using the Lsoda algorithm. Equilibrium properties were extracted after $1.5$ps simulations with the time step being determined by the Lsoda algorithm. The energy range of the lead was taken to range from $-0.1$ to $0.1$ except for the case of strong vibrational coupling where it taken to range from  $-0.2$ to $0.2$. $10000$ trajectories were found to be sufficient for convergence in all regimes tested.

\bibliography{EC}